\documentclass[twocolumn,showpacs,aps,prd,floatfix]{revtex4}
\usepackage{graphicx}
\usepackage{dcolumn}
\usepackage{amsmath}
\usepackage{epsfig}

\newcommand{\BABARPubYear}    {06}
\newcommand{\BABARPubNumber}  {033}
\newcommand{\SLACPubNumber} {11915}

\input pubboard/babarsym

\long\def\inst#1{\par\nobreak\kern 4pt\nobreak
    {\it #1}\par\vskip 10pt plus 3pt minus 3pt}

\graphicspath{%
{./figures/}%
{.}
}

\begin{document}

\def\Vubno  {\ensuremath{V_{ub}}\xspace}
\def\MKstarz  {\ensuremath{M_{\Kp\pim}}\xspace}
\def\MBz {\ensuremath{M_{\Bz}}\xspace}
\def\helicity {\ensuremath{\mathrm{\theta_{H}}}\xspace}
\def\thetaKS {\ensuremath{\theta_{\KS}}\xspace}
\def\thrust {\ensuremath{\theta_{\mathrm{T}}}\xspace}
\def\xmax {\ensuremath{x_{\mathrm{max.}}}\xspace}
\def\mkk {\ensuremath{m_{KK}}\xspace}
\def\mkpi {\ensuremath{m_{K\pi}}\xspace}
\def\Deltas {\ensuremath{\Delta S_{\phi\Kz}}\xspace}
\def\Deltasabs {\ensuremath{|\Delta S_{\phi\Kz}|}\xspace}
\def\xihat {\ensuremath{\hat{\xi}_{\phi\Kz} }\xspace}
\def\xiphikz {\ensuremath {\xi_{\phi\Kz} }\xspace}
\def\Dmass {\ensuremath {M_{\pi\KS} }\xspace}
\def\phimass {\ensuremath {M_{\Kp\Km} }\xspace}
\def\BzorBzb  {\ensuremath {\stackrel{_{(-)}}{B}{\kern-.25em}^{0}} \xspace}

\begin{flushleft}
\babar-PUB-\BABARPubYear/\BABARPubNumber \\
SLAC-PUB-\SLACPubNumber \\[6mm]
\end{flushleft}

\title{Search for the Decay of a $\Bz$ or $\Bzb$ Meson to $\Kstarzb\Kz$
or $\Kstarz\Kzb$
}

%
\author{B.~Aubert}
\author{R.~Barate}
\author{M.~Bona}
\author{D.~Boutigny}
\author{F.~Couderc}
\author{Y.~Karyotakis}
\author{J.~P.~Lees}
\author{V.~Poireau}
\author{V.~Tisserand}
\author{A.~Zghiche}
\affiliation{Laboratoire de Physique des Particules, F-74941 Annecy-le-Vieux, France }
% ...
\collaboration{The \babar\ Collaboration}
\noaffiliation

\begin{abstract}

\noindent
We present a search for the decay of a
$\Bz$ or $\Bzb$ meson to a $\Kstarzb\Kz$ 
or $\Kstarz\Kzb$ final state,
using a sample of approximately 
232~million \BB events collected with the {\babar} detector
at the \pep2 asymmetric energy \epem collider at SLAC.
The measured branching fraction is
$\BR(\Bz\ra\Kstarzb\Kz)+\BR(\Bz\ra\Kstarz\Kzb)=
(0.2^{+0.9}_{-0.8}\,^{+0.1}_{-0.3})\times 10^{-6}$.
We obtain the following upper limit 
for the branching fraction at 90\% confidence level:
$\BR(\Bz\ra\Kstarzb\Kz)+\BR(\Bz\ra\Kstarz\Kzb)
<1.9\times 10^{-6}$.
We use our result to constrain the Standard Model prediction
for the deviation of the CP asymmetry in 
$\Bz\rightarrow\phi\Kz$ from~$\stwob$.

\end{abstract}

\pacs{13.25.Hw, 12.15.Hh, 11.30.Er}
\maketitle

\vfill

\newpage

\section{INTRODUCTION}
\label{sec:Introduction}

\begin{figure}[t]    
\begin{center}
\includegraphics[width=6cm]{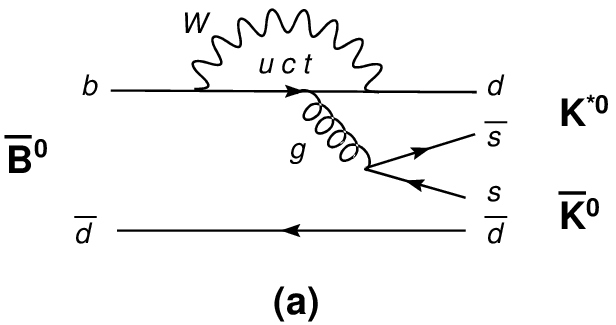} \\[4mm]
\includegraphics[width=6cm]{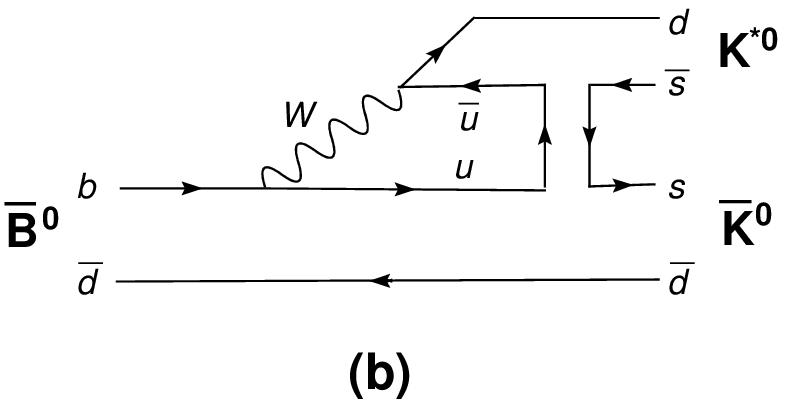}
\end{center}
\caption {
Feynman diagrams for $\Bzb \rightarrow \Kstarz \Kzb$:
(a)~penguin diagrams and
(b) $b\ra u$ rescattering diagram.
}
\label{fig-kstar-ks}
\end{figure}

This paper describes a search for the decay 
of a \Bz or \Bzb meson to a $\Kstarzb\Kz$ or $\Kstarz\Kzb$ final state.
Henceforth,
we use $\Bz\ra\Kstarzb\Kz$
to refer to both \Bz and \Bzb decays
and to the $\Kstarzb\Kz$ and $\Kstarz\Kzb$ decay channels.
In the Standard Model (SM),
$\Bz\ra\Kstarzb\Kz$ decays are described by 
$\b\rightarrow\d\ssbar$ diagrams
such as those shown in Fig.~\ref{fig-kstar-ks}.
Figure~\ref{fig-kstar-ks}(a) illustrates
$\b\ra\d$ ``penguin'' transitions.
A so-called rescattering process,
effectively a tree-level $\b\ra\d\u\ubar$ weak decay
followed by the long distance production of a $\s\sbar$ pair,
is shown in Fig.~\ref{fig-kstar-ks}(b).
Other rescattering diagrams,
e.g., with an intermediate \c quark loop rather than
a \u quark loop, are also possible.
Note that the rescattering diagrams can be considered to be 
the long distance components of the corresponding penguin diagrams,
in which the quark in the intermediate loop 
approaches its mass shell.

The SM prediction for the branching fraction of $\Bz\ra\Kstarzb\Kz$
is about \mbox{$0.5\times 10^{-6}$}~\cite{bib-kstar-ks-br}--\cite{bib-rpv-susy}.
Extensions to the SM can yield significantly larger 
branching fractions, however.
For example,
models incorporating supersymmetry with 
R-parity violating interactions 
predict branching fractions as large as about
\mbox{$8\times 10^{-6}$}~\cite{bib-rpv-susy}.
The event rates corresponding to this latter prediction are
well within present experimental sensitivity.
Currently, there are no experimental results for $\Bz\ra\Kstarzb\Kz$.
Searches for the related non-resonant decay
$\Bz\ra\Km\pip\Kz$ are reported in Ref.~\cite{bib-cleo-kp-pm-kz-2002}.

\begin{figure}[t]
\begin{center}
\includegraphics[width=6cm]{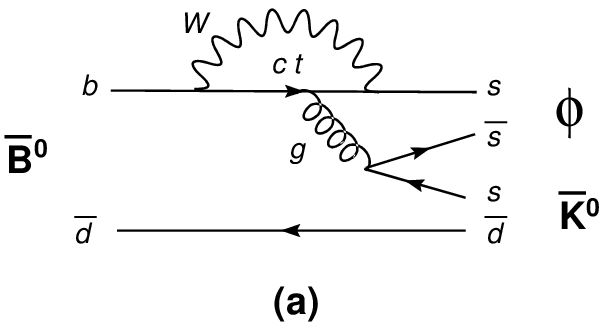} \\[4mm]
\includegraphics[width=6cm]{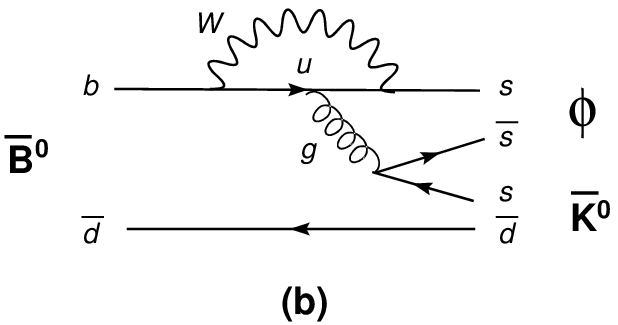} \\[4mm]
\includegraphics[width=6cm]{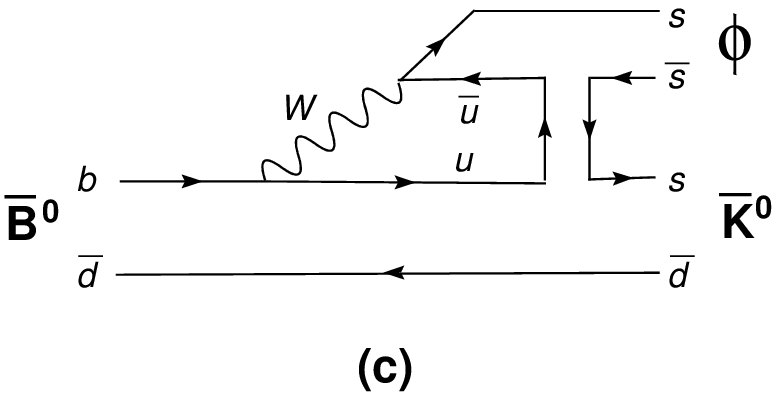}
\caption {
(a) CKM Dominant
and (b,c)~CKM suppressed diagrams for $\Bzb \rightarrow \phi\Kzb$.
}
\label{fig-phi-ks}
\end{center}
\end{figure}

At present, little experimental information is available
for $b\rightarrow d$ transitions.
Such processes can provide important tests of the 
quark-flavor sector of the SM as discussed,
for example, in Ref.~\cite{bib-fleisher-2005}.
Our study can also help to clarify
issues concerning potential differences between determinations of
$\stwob$ from tree- and penguin-dominated processes,
where $\beta$ is an angle of the Unitarity Triangle.
Such differences can provide a signal for physics 
beyond the SM~\cite{bib-new-physics-penguin}.
In particular,
our study is relevant for the interpretation of the
time dependent CP asymmetry
obtained from $\Bz\rightarrow\phi\Kz$ decays.
(For a review of the Unitarity Triangle
and \stwob measurements
based on $\Bz\rightarrow\phi\Kz$ decays, 
see Sec.~12 of Ref.~\cite{bib-pdg}.)
In the SM,
this decay is dominated by the
$\b\rightarrow\s$ penguin diagrams shown in Fig.~\ref{fig-phi-ks}(a).
In addition, 
sub-dominant SM processes with a different weak phase,
such as those shown in Figs.~\ref{fig-phi-ks}(b) and~(c)
involving the CKM matrix element \Vubno,
contribute at a level that 
is believed to be small~\cite{bib-ds-small}.
The deviation
of the CP asymmetry in $\Bz\rightarrow\phi\Kz$ decays
from $\stwob$ because of these sub-dominant processes
is referred to as~\Deltas.

Grossman {\it et al.}~\cite{gross-su3}
introduced a method to obtain a SM bound on \Deltas,
using SU(3) flavor symmetry to relate 
sub-dominant terms such as those shown
in Figs.~\ref{fig-phi-ks}(b) and~(c)
to the corresponding terms in strangeness-conserving
processes such as those shown in Fig.~\ref{fig-kstar-ks}.
To determine this bound,
measurements of the branching fractions of
11 \Bz decay channels are required
($\Kstarzb\Kz$, $\Kstarz\Kzb$,
and $hh^\prime$ with $h=\phi$, $\omega$ or $\rho^0$
and $h^\prime=\eta$, $\eta^\prime$ or~$\pi^0$). 
Experimental results are currently available
for all these channels except the two in our study:
\mbox{$\Kstarzb\Kz$} and $\Kstarz\Kzb$.
Our measurements will therefore enable this bound on \Deltas
to be determined for the first time.
Note that there are not statistically significant signals
for any of the nine channels for which results are
currently available.

Our results might also help to constrain predictions
for other charmless, strangeness-conserving decays
such as $\Bz\ra\rho\pi$,
in which a \ddbar or \uubar pair couples to the
gluon in Fig.~\ref{fig-kstar-ks}(a)
rather than a \ssbar pair
(see, e.g., Table~III of Ref.~\cite{bib-chiang}).

\section{THE \babar\ DETECTOR AND DATASET}
\label{sec:babar}

The data used in this analysis were collected with the \babar\ detector
at the \pep2\ asymmetric \epem storage ring. 
The data sample consists of an
integrated luminosity of 210~fb\inverse
recorded at the \FourS resonance with a
center-of-mass (CM) energy of $\sqrt{s}=10.58$~\gev,
corresponding to $(232\pm 2)\times 10^{6}$ \BB events.
A data sample of 21.6~fb\inverse
with a CM energy 40~\mev below the \FourS resonance
is used to study background contributions from
light quark \epem\ra\qqbar ($q=u, d, s$ or $c$) continuum events.

The \babar\ detector is described in detail 
elsewhere~\cite{bib-babar-detector}.
Charged particles are reconstructed using a five-layer
silicon vertex tracker (SVT) and a 40-layer drift chamber (DCH)
immersed in a 1.5~T magnetic field.
Charged pions and kaons are identified
(particle identification) with likelihoods
for particle hypotheses constructed from specific 
energy loss measurements in the SVT and DCH
and from Cherenkov radiation angles measured in the
detector of internally reflected Cherenkov light.
Photons are reconstructed in the electromagnetic calorimeter.
Muon and neutral hadron identification are performed with the
instrumented flux return.

Monte Carlo (MC) events are used to determine
signal and background characteristics, 
optimize selection criteria, and evaluate efficiencies.
\BzBzb and \BpBm events,
and continuum events,
are simulated with the 
\evtgen~\cite{bib-evtgen} and {\tt Jetset}~\cite{bib-jetset} 
event generators, respectively.
The effective integrated luminosity of the MC samples is
at least four times larger than that of the data
for the \BzBzb and \BpBm samples,
and about 1.5 times that of the data
for the continuum samples.
In addition,
separate samples of specific \BzBzb decay channels are studied 
for the purposes of background evaluation
(see, e.g., the channels mentioned in Sec.~\ref{sec-background}).
All MC samples include simulation of the {\babar} 
detector response~\cite{bib-geant4}.

\section{ANALYSIS METHOD}

\subsection{EVENT SELECTION}
\label{sec-selection}

$\Bz\ra\Kstarzb\Kz$ event candidates are selected
through identification of
$\Kstarz\rightarrow\Kp\pim$ and $\Kz\ra\KS\rightarrow\pip\pim$ decays.
Throughout this paper,
the charge conjugate channels are implied unless otherwise noted.
As the first step in the selection process,
we identify events with at least five charged tracks
and less than 20~\gev of total energy.
\KS~candidates are formed by combining all
oppositely charged pairs of tracks,
by fitting the two tracks to a common vertex,
and by requiring the pair to have a fitted invariant mass
within 0.025~\gevcc of the nominal \KS mass
assuming the two particles to be pions.
The \KS candidate is combined in a vertex fit
with two other oppositely charged tracks,
associated with the \Kstarz decay,
to form a \Bz candidate.
These latter two tracks are each required to have a distance of closest
approach to the \epem collision point of less than 1.5~cm
in the plane perpendicular to the beam axis
and 10~cm along the beam axis.
Of the two tracks associated with the \Kstarz decay,
one is required to be identified as a kaon
and the other as a pion using the particle identification.
Charged kaons are identified with an efficiency and purity
of about 80\% and~90\%, respectively,
averaged over momentum.
The corresponding values for charged pions are 90\% and~80\%.

Our study utilizes an extended maximum likelihood (ML) technique 
to determine the number of signal and background events
(Sec.~\ref{sec-ml-fit}).
The fitted experimental variables
are $\DeltaE$, \mes, and the mass of the \Kstarz candidate \MKstarz,
with $\DeltaE\equiv E^{*}_{B} - E^{*}_{beam}$ and
$\mes\equiv\sqrt{E^{*2}_{beam} - P^{*2}_{B}}$~\cite{bib-babar-detector},
where  $E^{*}_{B}$ and $P^*_{B}$ are the CM energy and momentum
of the \Bz candidate
and $E^{*}_{beam}$ is half the CM energy.
\MKstarz is determined by fitting the tracks from the
\Kstarz candidate to a common vertex.
We require events entering the ML fit to satisfy the following restrictions:
\begin{itemize}
  \item $|\DeltaE| < 0.15 \gev$,
  \item $5.2 < \mes < 5.3 \gevcc$,
  \item $ 0.72 < \MKstarz < 1.20 \gevcc$.
\end{itemize}
Note that virtually all well reconstructed signal events
satisfy these criteria.

We further impose the following criteria.
The selection values are optimized
to minimize the estimated upper limit on 
the $\Bz\ra\Kstarzb\Kz$ branching fraction
by comparing the number of expected signal~\cite{bib-chiang}
and background events as the selection values are changed.
\begin{itemize}
\item \Bz criteria: The $\chi^2$ probability of the fitted \Bz vertex 
  is required to exceed~0.003.
\item \Kstarz criteria: \Kstarz candidates are required to satisfy
  $|\cos\helicity | > 0.50 $, 
  where \helicity is the helicity angle in the \Kstarz rest frame,
  defined as the angle between the direction of the boost from the 
  \Bz rest frame and the \Kp momentum.
  
\item
  \KS criteria: The $\chi^2$ probability of the fitted \KS vertex 
  is required to exceed~$0.06$.
  The fitted \KS mass is required to lie within 10.5\mevcc of the peak of
  the reconstructed \KS mass distribution.
  (For purposes of comparison, 
  one standard deviation of the \KS mass resolution is about 3~\mevcc.)
  The \KS decay length significance, defined by the distance between
  the \Kstarz and \KS decay vertices 
  divided by the uncertainty on that quantity,
  is required to be larger than~$3$.
  The angle between the \KS flight direction and its momentum vector,
  $\thetaKS$,
  is required to satisfy $\cos\thetaKS>0.997$,
  where the \KS flight direction is defined by the direction
  between the \Kstarz and \KS decay vertices.
\item
  Event shape criteria: To separate signal events from the continuum background,
  we apply selection requirements on global momentum properties.
  \Bz~mesons in \FourS decays are produced almost at rest.
  Therefore, the \Bz decay products are
  essentially isotropic in the event~CM.
  In contrast,
  continuum \mbox{\epem\ra\qqbar} events at the \FourS energy
  are characterized by back-to-back two-jet-like event structures
  because of the relatively small masses of hadrons containing
  \u, \d, \s and \c quarks.
  As a means to separate signal from continuum background events,
  we calculate the Legendre polynomial-like terms $L_0$ and $L_2$
  defined by
  $L_0 = \sum_{\mathrm{r.o.e.}} p_i$ and
  $L_2 = \sum_{\mathrm{r.o.e.}} \frac{p_i}{2}\,(3\cos^2\theta_i-1)$,
  where $p_i$ is the magnitude of the 3-momentum of a particle 
  and $\theta_i$ is its polar angle with respect 
  to the thrust~\cite{bib-thrust} axis,
  with the latter determined
  using the candidate \Bz decay products only.
  These sums are performed over all particles in the event not 
  associated with the \Bz decay (``rest-of-event'' or r.o.e.).
  $L_0$ and $L_2$ are evaluated in the CM frame.
  We require $0.374 \, L_0 - 1.179\, L_2>0.15$.
  The coefficients of $L_0$ and $L_2$ are determined 
  with the Fisher discriminant method~\cite{bib-fisher}.
  To further reduce the continuum background,
  we also require $|\cos\thrust | < 0.55$,
  where \thrust is the angle between the momentum 
  of the \Bz candidate and the thrust axis,
  evaluated in the CM frame,
  with the thrust axis in this case determined 
  using all particles in the event {\it except} those associated 
  with the~\Bz candidate.
\end{itemize}
After applying the above criteria,
3.8\% of the selected events are found to contain 
more than one \Bz candidate.
For these events,
only the candidate with the largest \Bz vertex fit probability is retained.

Our selection procedure eliminates 99.78\% and 99.97\% of the 
\BB and continuum background MC events, respectively,
while retaining \mbox{$9.8\pm 0.1$\%} of the signal MC events.

\subsection{BACKGROUND EVALUATION}
\label{sec-background}

To identify residual backgrounds from $B$ decays,
we examine \BzBzb and \BpBm MC events that satisfy the
selection criteria of Sec.~\ref{sec-selection} and that
fall within the expected signal region of the \mes distribution,
defined by $5.271 < \mes < 5.286$~\gevcc.
The events so-identified are divided into four categories.
\begin{enumerate}
\item Events containing \Bz decays with the same 
  $\kaon\pi\pi\pi$ final state as the signal,
  such as $\Bz\rightarrow \Dmp \Kpm  \, (\Dmp\ra \pimp\KS)$,
  $\Bz\ra \Dmp \pipm ~(\Dmp \ra \Kpm \pimp \pimp)$,
  or $\Bz\rightarrow \Kpm \pimp \KS$.
  These channels are expected to peak in the signal regions of 
  \mes and \DeltaE but not in the signal region of~\MKstarz.
  The largest number of background events in this category arises from
  $\Bz\rightarrow \Dmp\Kpm  \, (\Dmp\ra \pimp\KS)$.
  To reduce the contributions of this channel,
  we apply a veto on the \mbox{$\pimp\KS$} mass \Dmass 
  based on the invariant mass of the \KS and the pion 
  used to reconstruct the~\Kstarz.  
  A veto with $1.813 < \Dmass < 1.925$~\gevcc 
  (corresponding to $\pm 7$~standard deviations of a 
  Gaussian fit to the \Dmass MC distribution)
  removes \mbox{$64\pm 1$\%} of the $\Dmp\Kpm$ background MC events
  but only $4.4\pm 0.6$\% of the signal MC events,
  where the uncertainties are statistical.
  Note that the reconstructed \Dmass distribution has
  non-Gaussian tails.
\item Events containing \Bz decays with a kaon misidentified as a pion, 
  such as 
  $~\Bz\rightarrow \phi \KS \, (\phi\ra K^+K^-)$ or
  $~\Bz\rightarrow f^0 \KS ~(f^0 \ra \Kp \Km)$.
  This category of
  background is expected to peak in the \mes signal region,
  but not in the \MKstarz signal region,
  and to exhibit a peak in \DeltaE that is negatively displaced 
  with respect to the signal peak centered at zero. 
  The largest number of events in this category arises
  from $\Bz\rightarrow \phi\KS \, (\phi\ra \Kp\Km)$.
  We apply a veto on the $\Kp\Km$ mass~\phimass
  assuming the pion candidate used to reconstruct 
  the \Kstarz to be a kaon.
  The veto requires $ 1.0098 <\phimass< 1.0280\gevcc$
  (corresponding to $\pm 2.5$ standard deviations of a 
  Gaussian fit to the $\phimass$ MC distribution).
  This selection requirement eliminates \mbox{$87\pm 1$\%}
  of the $\phi$\KS background MC events
  but only \mbox{$1.2\pm 0.3$\%}  of the signal MC events.
\item Events containing \Bz decays with a pion misidentified as a kaon,
  such as $\Bz \ra \Dpm \pimp \,(\Dpm \ra \pipm \KS)$
  or $\Bz\rightarrow \rho^0 \KS ~(\rho^0 \ra \pipm \pimp)$.
  This category of background peaks in the \mes signal region
  but not in the \MKstarz signal region
  and exhibits a peak in \DeltaE that is positively
  displaced from zero. 
\item All remaining \BzBzb and \BpBm MC events 
  that do not fall into the three categories listed above,
  such as
  $\Bz\rightarrow \Kstarz  \g ~(\Kstarz \ra \Kpm \pimp)$,
  $\Bz\rightarrow \Dmp\Kpm ~(\Dmp \ra \mun \numb \KS)$,
  or $\Bz\rightarrow \etapr\KS ~(\etapr\ra\rho^0 \g)$.
  These events are characterized both by particle misidentification
  and an exchange of tracks between the $\B$ and $\Bbar$ decays.
  This class of events does not peak in \DeltaE.
\end{enumerate}
Based on scaling to the experimental luminosity,
1.0~event (rounded to the nearest integer)
is expected for each of the first three categories,
and 54 events for the fourth category.

We also consider potential background from the following source.
\begin{enumerate}
\item[5.]
  Events with the same 
  $\kaon\pi\pi\pi$ final state as our signal
  but with a $\Kpm\pimp$ S-wave decay amplitude,
  either non-resonant or produced, e.g., through
  $\Bz\ra K_0^{*0}(1430) \KS \;\;( K_0^{*0}(1430) \ra \Kpm \pimp)$ decays.
  These channels are expected to peak in the signal regions of
  \mes and \DeltaE but not in the signal region of~\MKstarz.
\end{enumerate}
There are no experimental results for $\Bz\ra K_0^{*0}(1430)\KS$.
Studies~\cite{bib-belle-2005}
of $\Bp\ra\Kp\pip\pim$ found a substantial
$\Bp\ra K_0^{*0}(1430)\pip$ resonant component, however.
To evaluate this potential source of background,
we generate 
$\Bz \ra K_0^{*0}(1430)\KS \;\;(K_0^{*0}(1430)\ra\Kp\pim)$
MC events.
After applying the criteria described in Sec.~\ref{sec-selection},
only \mbox{$1.4\pm 0.1$}\% of these events remain.
More importantly,
the interference between the $K^{*0}(890)$ and 
\mbox{S-wave} $K\pi$ amplitudes is expected to cancel
if the detection efficiency is symmetric in 
the candidate \Kstarz $\cos\helicity$ distribution.
Through MC study,
we verify that our efficiency is symmetric in $\cos\helicity$
to better than about~$10$\%.
This allows us to treat potential S-wave $\Kpm\pimp$ background
as an independent component in the ML fit.

\subsection{FIT PROCEDURE}
\label{sec-ml-fit}

An unbinned extended maximum likelihood fit is used 
to determine the number of signal and background events in the data.
The extended likelihood function ${\cal L}$ is defined by
\begin{equation}
  {\cal L} = \exp{\left( -\sum_{i=1}^7 n_i\right) } 
      \prod_{j=1}^N \left[\sum_{i=1}^{7} n_i {\cal P}_i \right],
\end{equation}
where $N$ is the number of observed events
and $n_i$ are the yields of the seven event categories:
signal, continuum background,
and the five \BB background categories from Sec.~\ref{sec-background}.
The correlations between the three fitted observables are found to be small
($\lsim 10$\% in both signal MC and background).
Therefore, we define the functions ${\cal P}_i$ to be 
products of three independent probability density functions (PDFs),
one for each of \DeltaE, \mes, and \MKstarz.
We account for effects related to residual correlations between the variables
through the bias correction and evaluation of systematic uncertainties
discussed in Secs.~\ref{sec-results} and~\ref{sec:Systematics}.

The signal PDFs are defined by a double Gaussian 
distribution for \DeltaE,
a Crystal Ball function~\cite{bib-crystal-ball} for \mes,
and a Breit-Wigner function for~\MKstarz. 
The parameters are fixed to values found from fitting signal MC events.
We verify that the signal MC predictions for the \DeltaE and \mes
distributions agree with the measured results from
$\Bz\ra\phi\KS$ decays~\cite{bib-babar-phiks}
to within the experimental statistical uncertainties.
The $\phi\KS$ channel is chosen for this purpose
because of its similarity to the $\Kstarzb\KS$ channel.

Separate PDFs are determined for the continuum background
and all five categories of \BB background itemized in Sec.~\ref{sec-background}.
The background PDFs are defined by combinations of polynomial,
Gaussian, \mbox{ARGUS~\cite{bib-argus},} 
and Breit-Wigner functions fitted to MC events,
with the exception of the PDFs for the S-wave $\Kpm\pimp$ component
for which the \DeltaE and \mes PDFs are set equal to those of the signal
while the \MKstarz PDF is based on the scalar $K\pi$
lineshape determined by the LASS Collaboration~\cite{bib-lass}.
All the fits of PDFs to MC distributions
yield values of $\chi^2$ per degree-of-freedom near unity.

The event yields of the continuum
and last two categories of \BB background from Sec.~\ref{sec-background}
are allowed to vary in the fits,
while those of the first three categories of \BB background are
set equal to the expected numbers given in Sec.~\ref{sec-background}.
The PDF shape parameters of the continuum events
are allowed to vary in the fit,
while those of the five \BB background categories are fixed.

\section{RESULTS}
\label{sec-results}

We find 682 data events that satisfy the selection criteria.
Application of the ML fit to this sample
yields $1.0^{+4.7}_{-3.9}$ signal events and
$660\pm 75$ continuum events
where the uncertainties are statistical.
These results and those for the 
\BB background yields are given in Table~\ref{table-result}.
Based on the SM branching fraction predictions of Ref.~\cite{bib-chiang},
5~signal events (rounded to the nearest integer) are expected.
The number of expected continuum events is~619.
The statistical uncertainty of the signal yield is defined by the change in the 
number of events required to increase the quantity
$-2\ln{\cal L}$ by one~unit from its minimum value,
and similarly for the other yields.
The statistical significance of the result,
defined by the square root of the difference
between the value of $-2\ln{\cal L}$ for zero signal events and at its minimum,
expressed in units of the statistical uncertainty,
is~0.28.

\begin{figure}[tp]
\begin{center}
\includegraphics[scale=.45]{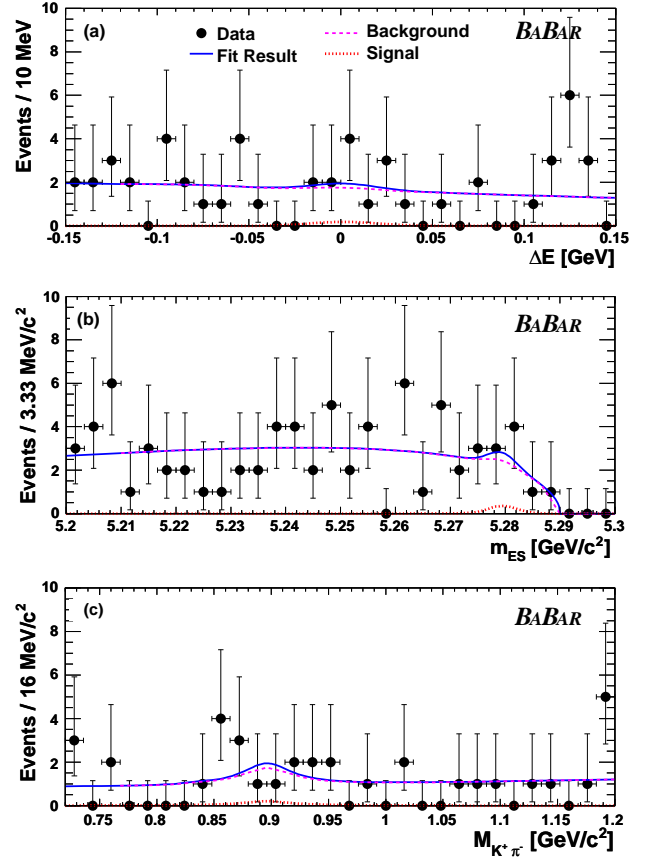}
\caption {
Distributions of \DeltaE, \mes, and \MKstarz.
The points with uncertainties show the data.
The curves show projections of the ML fit.
A selection requirement on the likelihood ratio 
has been applied as described in the text.
The solid curve shows the sum of all fitted components,
including the signal.
The dashed curve shows the sum of all background components.
The dotted curve (barely visible) shows the signal component.
}
\label{fig-projPlotUB}
\end{center}
\end{figure}

Figure~\ref{fig-projPlotUB} shows distributions
for each of the fitted variables.
To enhance the visibility of a potential signal,
events in Fig.~\ref{fig-projPlotUB} are required to satisfy 
${\cal L}_i(S)/[{\cal L}_i(S)+{\cal L}_i(B)]>0.6$,
where ${\cal L}_i(S)$ is the likelihood function for signal events
excluding the PDF of the plotted variable $i=\DeltaE,$ \mes or~\MKstarz,
and ${\cal L}_i(B)$ is the corresponding term
for all background components added together.
The points with uncertainties show the data.
The curves show projections of the ML fit
with the likelihood ratio restriction imposed.

\begin{table}[tp]
\begin{center}
\caption{Results from the maximum likelihood fit.
\BB~background categories 4 and 5 refer to the last two categories 
of background itemized in Sec.~\ref{sec-background}.
The yields for the first three \BB background categories
in Sec.~\ref{sec-background} are fixed to the 
estimated values of 1.0~event each.
The uncertainties on the yields, fit bias,
and efficiencies are statistical.
}
\vspace{3mm}
\begin{tabular} {l c}
\hline
Parameter   			& Value  \\
\hline
    &     \\
Number of events		&  682   \\
    &      \\
Signal yield 			& $1.0^{+4.7}_{-3.9}$	\\
          \\
Continuum background yield   	& $660\pm 75$	\\
	&   \\
\BB background category 4 yield & $17^{+74}_{-71}$    \\
    &        \\
\BB background category 5 yield & $1.4^{+6.4}_{-5.3}$    \\
     &    \\
ML fit bias (signal bias)	&  $-0.2\pm 0.3$  \\
        &        \\
MC signal efficiency    	& $9.8\pm 0.1$\%      \\
(including $\Dmp$ and $\phi$ mass vetos) & \\
    &     \\
Efficiency corrections  	&	  \\
\hspace{5mm}\KS tracking 	& 97.8\%  \\
\hspace{5mm}\Kstarz tracking    & 99.0\%  \\ 
\hspace{5mm}Final-state branching fractions   & 23.0\% \\  
  &       \\
Overall detection efficiency	& $2.2\pm0.1$\%     \\
   &   \\
\hline
   &   \\
\BR($\Bz\ra\Kstarzb\Kz)$
               & $(0.2^{+0.9}_{-0.8}\,^{+0.1}_{-0.3})\times 10^{-6}$    \\
\hspace*{5mm} $+\BR(\Bz\ra\Kstarz\Kzb)$   &  \\
   &     \\
Significance with systematics ($\sigma$)  &   0.26     \\
   &     \\
90\% CL upper limit on          & $<1.9\times 10^{-6}$    \\
\BR($\Bz\ra\Kstarzb\Kz)+\BR(\Bz\ra\Kstarz\Kzb$) &    \\[2mm]
\hline
\end{tabular}
\label{table-result}
\end{center}
\end{table}

We evaluate potential bias in the fitted signal yield
by applying the ML fit to 250 simulated data
samples constructed as described below.
The number of continuum background events in each sample 
is derived from a Poisson distribution, 
with a mean set equal to the number of continuum events found 
in the data, i.e., 660 events.
We generate \DeltaE, \mes, and \MKstarz continuum distributions
for each sample by randomly sampling the continuum PDFs using
the appropriate number of events for each sample.
The number of \BB background events in each sample is
determined in the analogous manner for each of the five \BB 
background categories separately.
For the first four categories of \BB background 
(all but the scalar $K\pi$ component),
the \DeltaE, \mes, and \MKstarz distributions
are generated by randomly selecting the appropriate number of
events from the corresponding MC sample.
For the scalar $K\pi$ component,
the distributions are generated by sampling the PDFs.

The number of signal events in each simulated sample is likewise 
determined from a Poisson distribution,
with a mean $N_{sig}^{P}$
initially set equal to the fitted signal yield
$N_{sig}=1.0$.
The signal \DeltaE, \mes, and \MKstarz distributions are
generated by randomly selecting the appropriate
number of signal MC events for each sample.
$N_{sig}^{P}$ is then adjusted until the mean signal yield 
from the 250 samples equals $N_{sig}$.
The ML fit bias is defined by 
$N_{bias}=N_{sig}-N_{sig}^{P}$
and is determined to be $-0.2\pm 0.3$~(stat.) events.
Therefore,
the corrected signal yield is $N_{sig}-N_{bias}=1.2$~events.

In our study, 
we can distinguish
$\Kstarzb\Kz$ from $\Kstarz\Kzb$ events
from the sign of the electric charge of the $K^\pm$.
However, we do not know the
flavor of the $B$ meson (\Bz or \Bzb) at decay.
Therefore, the observed signal yield is related to
the sum of the $\Bz\rightarrow \Kstarzb\Kz$ and
$\Bz\rightarrow \Kstarz\Kzb$
branching fractions through
\begin{equation}
  \label{eq-BR}
  \BR(\Bz\rightarrow \Kstarzb\Kz)
  + \BR(\Bz\rightarrow \Kstarz\Kzb)
    = \frac{N_{sig}-N_{bias}}{\epsilon\, N_{\BB}},
\end{equation}
where $\epsilon$ is the overall detection efficiency,
given by the product of the MC signal efficiency 
and three efficiency corrections (Table~\ref{table-result}).
The \KS and \Kstarz tracking corrections account for
discrepancies between the data and MC simulation,
while the correction for final-state branching fractions
accounts for the $\Kz\ra\KS$, $\KS\ra\pip\pim$, 
and $\Kstarz\ra\Kp\pim$ branching fractions,
which are not incorporated into the simulated
signal event sample.
The overall efficiency is $\epsilon=2.2$\%.
The factor $N_{\BB}$ in Eq.~(\ref{eq-BR})
is the number of \BB events in the initial data sample
of~210~fb\inverse.
We assume equal decay rates of the \FourS to \BzBzb and~\BpBm.

We find the sum of the  branching fractions to be
$\BR(\Bz\rightarrow \Kstarzb\Kz)
+ \BR(\Bz\rightarrow \Kstarz\Kzb)
=(0.2^{+0.9}_{-0.8}\,^{+0.1}_{-0.3})\times 10^{-6}$,
where the first uncertainty is statistical and
the second is systematic.
The systematic uncertainty is discussed in Sec.~\ref{sec:Systematics}.
We determine a Bayesian 90\% confidence level (CL) 
upper limit assuming a uniform prior probability distribution.
First,
the likelihood function is modified to incorporate systematic uncertainties
through convolution with a Gaussian distribution whose
standard deviation is set equal to the total systematic uncertainty.
The 90\% CL upper limit is then defined to be the value of
the branching fraction below which lies 90\% of the
total of the integral of the modified likelihood function
in the positive branching fraction region.
We obtain
\mbox{$\BR(\Bz\rightarrow\Kstarzb\Kz)$}$\,+\,$
\mbox{$\BR(\Bz\rightarrow\Kstarz\Kzb)$}$\,<1.9\times 10^{-6}$.
We also use the modified likelihood function to determine the
significance of our branching fraction result including systematics.
This result is listed in Table~\ref{table-result}.

\section{SYSTEMATIC UNCERTAINTIES}
\label{sec:Systematics}

To evaluate systematic uncertainties,
we consider effects associated with the ML fit,
the \BB background estimates,
the efficiency corrections,
the total number of \BB events,
and the $\KS\ra\pip\pim$ branching fraction.
Table~\ref{table:sys-summary} provides a summary.

To estimate the systematic uncertainty related to the signal PDFs,
we independently vary the 11 parameters used to characterize the
signal \DeltaE, \mes, and \MKstarz PDFs.
The mean and standard deviation of the central \DeltaE Gaussian distribution,
and the mean of the \mes Crystal Ball function,
are varied by the statistical uncertainties found by fitting
the corresponding quantities to data in a recent study of
$\Bz\ra\phi\Kz$ decays~\cite{bib-babar-phiks}.
We vary the standard deviation of the \mes Crystal Ball function
to account for observed variations between different run periods.
The width of the \MKstarz Breit-Wigner function is varied by $\pm 0.01$~\gevcc.
The remaining six signal PDF parameters are varied by one standard
deviation of their statistical uncertainties found in the
fits to the MC distributions (Sec.~\ref{sec-ml-fit}),
taking into account correlations between parameters.
For variations of all 11 parameters,
the percentage change in the signal yield compared to the standard fit is taken 
as that parameter's contribution to the overall uncertainty.  
The total systematic uncertainty associated with the
signal PDFs is obtained by adding these 11 contributions in quadrature.
The largest contributions are from the variations of the \DeltaE mean and
standard deviation (about 0.3 signal events each).

The systematic uncertainty attributed to the fit bias
is defined by adding two terms in quadrature.
The first term is the statistical uncertainty of this bias (Table~\ref{table-result}).
The second term is defined by changing the method used to
determine the bias.
Specifically,
we evaluate this bias by generating the \DeltaE, \mes, and \MKstarz distributions
of the fourth \BB background category in Sec.~\ref{sec-background}
using the PDFs rather than sampling MC events,
for the 250 simulated data samples:
the difference between the results of this method and the standard one allows us
to assess the effect of residual correlations between the variables.
The fourth category of \BB background events is chosen
because it dominates the \BB background.
The difference between the corrected mean signal yield 
and the standard result defines the second term.

To estimate an uncertainty associated with the \BB background,
we vary the assumed numbers of events for the
three \BB background categories for which these numbers are fixed,
i.e., the first three background categories of Sec.~\ref{sec-background}.
Specifically,
we independently vary these numbers by $+2$ and $-1$ events
from their standard values of unity,
and determine the quadrature sum of
the resulting changes in the signal yield.

A systematic uncertainty associated with the presumed scalar $K\pi$ lineshape
is defined by the difference between the signal yield
found using the LASS lineshape
and a uniform (i.e., flat) $K\pi$ mass distribution.

Systematic uncertainties for the \KS reconstruction efficiency,
and for the tracking and particle identification efficiencies of the 
\Kp and \pim used to reconstruct the \Kstarz,
account for known discrepancies between the data and MC simulation
for these quantities.
Similarly,
the MC simulation overestimates the number of selected events
compared to data for values of \mbox{$|\cos\thrust |$} less than about~0.9.
We assign a 5\% systematic uncertainty to account for this effect.

The systematic uncertainty associated with the number of \BB pairs
is determined to be 1.1\%.
The uncertainty in the $\KS\ra\pip\pim$ branching fraction
is taken from Ref.~\cite{bib-pdg}.

The total systematic uncertainty is defined by
adding the above-described items in quadrature.

\begin{table}[tp]
\begin{center}
\caption {Summary of systematic uncertainties.}
\vspace{3mm}
\begin{tabular} {l c}
\hline
Systematic effect               & Uncertainty  \\
\hline
                      &  \\
ML fit procedure (events)		   &	\\
\hspace{5mm} Signal PDF parameters   &  0.5  	\\
\hspace{5mm} Fit bias	             &  0.5  	\\
\hspace{5mm} \BB background yields   &  0.1 	\\
                      &           \\
Total uncertainty from ML fit (events)   &  0.7     \\
                      &  \\
\hline
                      &  \\
Scalar $K\pi$ lineshape (events)     & $^{+0.0}_{-1.4}$  \\
           &     \\
\hline
                      &  \\
Efficiency corrections (\%)            &        \\
\hspace{5mm} \KS reconstruction    &  1.4\%  	\\
\hspace{5mm} \Kstarz tracking      &  2.8\%  	\\
\hspace{5mm} \Kstarz Particle identification efficiency &  0.8\%  	\\
$\cos\thrust$ selection requirement                &  5.0\%     \\
Number of \BB pairs                &  1.1\%	\\  
\BR ($\KS \ra \pipm \pimp$)	   &  0.1\%	\\
             &          \\
Total uncertainty from corrections       &  6.1\%     \\
                      &  \\
\hline
       &      \\
Total systematic uncertainty for \BR($\times 10^{6}$) &  
    {\large{$^{+0.1}_{-0.3}$}}     \\
	&         \\
\hline
\end{tabular}
\label{table:sys-summary}
\end{center}
\end{table}

\section{SUMMARY AND DISCUSSION}
\label{sec-Summary}

In this paper,
we present the first experimental results for
the decay $\Bz(\Bzb)\rightarrow\Kstarzb\Kz$.
From a sample of about 232 million \BB events,
we observe $1.0^{+4.7}_{-3.9}$ $\Bz\rightarrow\Kstarzb\Kz$ event candidates.
The corresponding measured sum of branching fractions is
$\BR(\Bz\rightarrow\Kstarzb\Kz)
+ \BR(\Bz\rightarrow\Kstarz\Kzb)
=(0.2^{+0.9}_{-0.8}\,^{+0.1}_{-0.3})\times 10^{-6}$.
We obtain a 90\% confidence level upper limit of
$\BR(\Bz\rightarrow\Kstarzb\Kz)
+\BR(\Bz\rightarrow\Kstarz\Kzb)
<1.9\times 10^{-6}$.
This result constrains certain extensions of the SM,
such as the R-parity violating supersymmetry models
described in Ref.~\cite{bib-rpv-susy}.

Our result also can be used to determine an upper bound on \Deltas,
as mentioned in the introduction.
The amplitude~$A$ for $\Bz\ra\phi\Kz$ can be expressed as~\cite{gross-su3}
\begin{equation}
    A=V_{cb}^{*} V_{cs} a^c + V_{ub}^{*} V_{us} a^u,
  \label{eq-amplitude}
\end{equation}
with $a^c=p^c-p^t$ and $a^u=p^u-p^t$,
where $p^i$ is the hadronic amplitude of the penguin diagram
with intermediate quark $i=\u$, \c or \t
[see Figs.~\ref{fig-phi-ks}(a) and~(b)].
The CKM factor multiplying $a^u$ in Eq.~(\ref{eq-amplitude})
is suppressed by ${\cal O}(\lambda^2)$
relative to the factor multiplying $a^c$,
where $\lambda=0.224$~\cite{bib-pdg} is the sine of the Cabibbo angle.
Therefore,
the diagrams in Fig.~\ref{fig-phi-ks}(a)
are expected to dominate $\Bz\ra\phi\Kz$ decays.
As described in Ref.~\cite{gross-su3}, 
\Deltas is given by
\begin{equation}
   \label{eqn:deltaS}
   \Deltas 
     = 2 \cos 2\beta \, \sin \gamma \, \cos \delta
        \, \left| \xiphikz \right|,
\end{equation}
with
\begin{equation}
  \xiphikz \equiv \frac{ V^*_{ub} V_{us} \ a^u }
                      { V^*_{cb} V_{cs} \ a^c },
\end{equation}
where $\delta$ and $\gamma$ are the strong and weak
phase differences, respectively, 
between $a^u$ and $a^c$.

Analogous to Eq.~(\ref{eq-amplitude}),
the amplitude~$A^\prime$
for $\Bz\ra\Kstarzb\Kz$ 
can be expressed as~\cite{gross-su3}
\begin{equation}
    A^\prime=V_{cb}^{*} V_{cd} b^c + V_{ub}^{*} V_{ud} b^u.
  \label{eq-kkstar-amplitude}
\end{equation}
In contrast to Eq.~(\ref{eq-amplitude}),
neither term in Eq.~(\ref{eq-kkstar-amplitude})
is suppressed by CKM factors relative to the other.
As an effective tree-level process,
it is therefore possible that the diagram of Fig.~\ref{fig-kstar-ks}(b)
dominates $\Bz\ra\Kstarzb\Kz$ decays.
(This assumption yields the most conservative limit on \Deltas.)

The method of Grossman~{\it et al.}~\cite{gross-su3}
consists of using SU(3) flavor symmetry to relate $b^c$ and $b^u$ 
in Eq.~(\ref{eq-kkstar-amplitude})
to $a^c$ and $a^u$ in Eq.~(\ref{eq-amplitude})
to obtain a bound on the quantity \xihat defined by
\begin{equation}
  \xihat \equiv\frac{V_{us}}{V_{ud}}
        \left(\frac{V_{cb}^* V_{cd}\, a^c + V_{ub}^*V_{ud} \, a^u }{A}\right),
\end{equation}
with $A$ given by Eq.~(\ref{eq-amplitude}).
The bound on \xihat is derived using the branching
fractions of 11 strangeness-conserving charmless \Bz decays:
\begin{eqnarray}
  \left| \xihat \right| & \leq & \left| \frac{V_{us}}{V_{ud}}\right|
       \left\{
       0.5 \,  \sqrt{ \frac{ 2 \, \left[
             {\cal B}(\Kstarzb\Kz)
           + {\cal B}(\Kstarz\Kzb)\right] }{{\cal B}(\phi K^0)} } \right. 
            \nonumber \\
      & &  \left.
       + \sum_{i=1}^9 C_i \ \sqrt{ \frac{{\cal B}(f_i)}{{\cal B}(\phi K^0)} }
        \right\},
  \label{eq-xi}
\end{eqnarray}
where the $C_i$ are SU(3) coefficients
and where the nine final states $f_i=hh^\prime$ 
are specified in the introduction.
\xihat is related to \xiphikz through~\cite{gross-su3,bib-engelhard-su3}
\begin{equation}
|\xihat|^2 
  = \frac{ 
       \left\vert\frac{V_{us}V_{cd}}{V_{cs}V_{ud}} \right\vert^2
       + \left\vert \xiphikz\right\vert^2
       + 2\cos\gamma\, Re\left( \frac{V_{us}V_{cd}}{V_{cs}V_{ud}}\,\xiphikz \right)
     }{
       1 + \left\vert \xiphikz \right\vert^2 + 2\cos\gamma\, Re\left(\xiphikz \right)
     }.
  \label{eq-xi-xihat}
\end{equation}
The observed rates of strangeness-conserving processes,
potentially dominated by $b\ra\u$ rescattering transitions
such as are illustrated in Fig.~\ref{fig-kstar-ks}(b),
are therefore
used to set limits on the contributions of the SM-suppressed 
$b\ra\u$ terms shown in Figs.~\ref{fig-phi-ks}(b) and~(c), 
i.e., to set limits on transitions
which cause a deviation of the CP asymmetry 
in $\Bz\rightarrow\phi\Kz$ decays from~$\stwob$.

We evaluate a 90\% CL upper limit on \Deltasabs
by generating hypothetical sets of
branching fractions for the 11 required SU(3)-related decays: 
$\Kstarz\Kzb$, $\Kstarzb\Kz$, and $hh^\prime$.
Branching fraction values are chosen using bifurcated
Gaussian probability distribution functions
with means and bifurcated widths
set equal to the measured branching fractions and asymmetric uncertainties.
For the measurements of the branching fractions
of the nine channels not included in the
present study, see Refs.~\cite{bib-9-measurements,bib-rho0-pi0}.
Negative generated branching fractions are discarded.
For each set of hypothetical branching fractions,
we compute a bound on \Deltasabs
using Eqs.~(\ref{eqn:deltaS}) and~(\ref{eq-xi}).
For the unknown phase term $\cos\delta$ in Eq.~(\ref{eqn:deltaS}),
we sample a uniform distribution between $-1$ and~$1$.
Similarly,
the weak phase angle $\gamma$ is chosen by selecting
values from a uniform distribution between 38 and 79 degrees,
corresponding to the 95\% confidence level interval for $\gamma$ 
given in Ref.~\cite{bib-ckm-fitter-2005}.
(A flat distribution is chosen for $\gamma$ because the 
likelihood curve in Ref.~\cite{bib-ckm-fitter-2005} is non-Gaussian.)
For $\beta$, 
we use~$\stwob=0.687$~\cite{bib-rho0-pi0}.
For each iteration of variables,
Eq.~(\ref{eq-xi-xihat}) is solved numerically for $|\xiphikz|$.

We find that 90\% of the hypothetical \Deltasabs bounds lie below~0.42.
Our study thus allows the SU(3) bound from Ref.~\cite{gross-su3},
viz.,~$\Deltasabs<0.42$ at 90\% CL,
to be determined for the first time.
To assess the contribution of the $\Kstarzb\Kz$ channel on this result,
we repeat the procedure described in the previous paragraph with
the $\Bz\rightarrow\Kstarzb\Kz$ branching fraction and uncertainties
set to zero:
the corresponding result is~0.32.
Potential future measurements
of $\Bz\rightarrow\Kstarzb\Kz$ yielding a significantly smaller UL
and uncertainties would therefore have a significant impact 
on the \Deltasabs bound.
As a cross check,
we also determine the SU(3) bound assuming the weak phase angle $\gamma$ 
to be distributed according 
to a Gaussian distribution with a mean of $58.5^\circ$ 
and a standard deviation of 5.8$^\circ$~\cite{bib-ut-fit-group}:
this yields $\Deltasabs<0.43$ at 90\%~CL.
Our analysis does not account for SU(3) flavor breaking effects,
generally expected to be on the order of~30\%.
However,
the method is conservative in that it assumes all hadronic
amplitudes interfere constructively.

\section{ACKNOWLEDGMENTS}
\label{sec:Acknowledgments}

We are grateful for the 
extraordinary contributions of our \pep2\ colleagues in
achieving the excellent luminosity and machine conditions
that have made this work possible.
The success of this project also relies critically on the 
expertise and dedication of the computing organizations that 
support \babar.
The collaborating institutions wish to thank 
SLAC for its support and the kind hospitality extended to them. 
This work is supported by the
US Department of Energy
and National Science Foundation, the
Natural Sciences and Engineering Research Council (Canada),
Institute of High Energy Physics (China), the
Commissariat \`a l'Energie Atomique and
Institut National de Physique Nucl\'eaire et de Physique des Particules
(France), the
Bundesministerium f\"ur Bildung und Forschung and
Deutsche Forschungsgemeinschaft
(Germany), the
Istituto Nazionale di Fisica Nucleare (Italy),
the Foundation for Fundamental Research on Matter (The Netherlands),
the Research Council of Norway, the
Ministry of Science and Technology of the Russian Federation, and the
Particle Physics and Astronomy Research Council (United Kingdom). 
Individuals have received support from 
CONACyT (Mexico), the Marie-Curie Intra European Fellowship program (European Union),
the A. P. Sloan Foundation, 
the Research Corporation,
and the Alexander von Humboldt Foundation.

\end{document}